\documentclass{article}

\usepackage{arxiv}

\usepackage[utf8]{inputenc} 
\usepackage[T1]{fontenc}    
\usepackage{hyperref}       
\usepackage{url}            
\usepackage{booktabs}       
\usepackage{amsfonts}       
\usepackage{nicefrac}       
\usepackage{microtype}      
\usepackage{lipsum}		
\usepackage{graphicx}
\usepackage{natbib}
\usepackage{doi}

\usepackage{xcolor}
\usepackage{caption}
\usepackage{subcaption}     
\usepackage{amsmath}
\usepackage{lscape}

\title{A machine-learning sleep-wake classification model using a reduced number of features derived from photoplethysmography and activity signals}

\author{Douglas A.~Almeida \\
	Heart Institute (InCor)\\
	Clinics Hospital University of \\
 Sao Paulo Medical School\\
	Sao Paulo - SP - Brazil \\
	\texttt{douglas.andrade@hc.fm.usp.br} \\
	\And
	Felipe M.~Dias \\
	Heart Institute (InCor)\\
	Clinics Hospital University of \\
 Sao Paulo Medical School\\
	Sao Paulo - SP - Brazil \\
	\texttt{f.dias@hc.fm.usp} \\
 \And
	Marcelo A.~F.~Toledo \\
	Heart Institute (InCor)\\
	Clinics Hospital University of \\
 Sao Paulo Medical School\\
	Sao Paulo - SP - Brazil \\
	\texttt{marcelo.arruda@hc.fm.usp.br} \\
 \And
	Diego A.~C.~Cardenas \\
	Heart Institute (InCor)\\
	Clinics Hospital University of \\
 Sao Paulo Medical School\\
	Sao Paulo - SP - Brazil \\
	\texttt{diego.cardona@hc.fm.usp.br} \\
 \And
	Filipe A.~C.~Oliveira \\
	Heart Institute (InCor)\\
	Clinics Hospital University of \\
 Sao Paulo Medical School\\
	Sao Paulo - SP - Brazil \\
	\texttt{filipe.acoliveira@hc.fm.usp.br} \\
 \And
	Estela ~Ribeiro \\
	Heart Institute (InCor)\\
	Clinics Hospital University of \\
 Sao Paulo Medical School\\
	Sao Paulo - SP - Brazil \\
	\texttt{estela.ribeiro@hc.fm.usp.br} \\
 \And
	Jose E.~Krieger \\
	Heart Institute (InCor)\\
	Clinics Hospital University of \\
 Sao Paulo Medical School\\
	Sao Paulo - SP - Brazil \\
	\texttt{j.krieger@hc.fm.usp.br} \\
  \And
	Marco A.~Gutierrez \\
	Heart Institute (InCor)\\
	Clinics Hospital University of \\
 Sao Paulo Medical School\\
	Sao Paulo - SP - Brazil \\
	\texttt{marco.gutierrez@incor.usp.br}
}

\date{}

\renewcommand{\shorttitle}{Sleep-wake classification}

\hypersetup{
pdftitle={A template for the arxiv style},
pdfsubject={q-bio.NC, q-bio.QM},
pdfauthor={David S.~Hippocampus, Elias D.~Striatum},
pdfkeywords={First keyword, Second keyword, More},
}

\begin{document}
\maketitle

\begin{abstract}
Sleep is a crucial aspect of our overall health and well-being. It plays a vital role in regulating our mental and physical health, impacting our mood, memory, and cognitive function to our physical resilience and immune system. The classification of sleep stages is a mandatory step to assess sleep quality, providing the metrics to estimate the quality of sleep and how well our body is functioning during this essential period of rest. Photoplethysmography (PPG) has been demonstrated to be an effective signal for sleep stage inference, meaning it can be used on its own or in a combination with others signals to determine sleep stage. This information is valuable in identifying potential sleep issues and developing strategies to improve sleep quality and overall health. In this work, we present a machine learning sleep-wake classification model based on the eXtreme Gradient Boosting (XGBoost) algorithm and features extracted from PPG signal and activity counts. The performance of our method was comparable to current state-of-the-art methods with a Sensitivity of 91.15 $\pm$ 1.16\%, Specificity of 53.66 $\pm$ 1.12\%, F1-score of 83.88 $\pm$ 0.56\%, and Kappa of 48.0 $\pm$ 0.86\%. Our method offers a significant improvement over other approaches as it uses a reduced number of features, making it suitable for implementation in wearable devices that have limited computational power.
\end{abstract}


\section{Introduction}
Sleep plays a vital role in maintaining good health for individuals of all ages \cite{ramar2021sleep}.
Inadequate sleep of poor quality or duration can lead to a host of chronic health problems, including cardiovascular diseases, diabetes, and obesity \cite{hoevenaar2011sleep} \cite{knutson2008associations} \cite{rahe2015associations}.
Moreover, sleep disorders are on the rise and have been linked to a decrease in Gross Domestic Product (GDP) in some countries \cite{hillman2006economic} \cite{streatfeild2021social}. These highlight the importance of ensuring that individuals get adequate and high-quality sleep to maintain their health and well-being. 

Polysomnography (PSG) is the gold standard exam for evaluating human sleep \cite{KRYSTAL2008}. The exam involves the simultaneous recording of multiple electrophysiological signals during sleep, such as Electrocardiogram (ECG), Photoplethysmography (PPG), Electroencephalogram (EEG), Electromyogram (EMG), etc. \cite{rundo2019polysomnography}. After registration, a specialist analyzes the recorded signals and labels different sleep stages based on time windows of 30 s. 
Nevertheless, PSG is an expensive and time-consuming procedure and can also suffer from labeling errors by the specialist. Additionally, PSG exams are performed in an unfamiliar environment, with multiple electrodes attached, which can affect sleep quality. 

With the rise of wearable devices, it is now possible to track several physiological signals, such as heart rate and blood oxygen levels, more cost-effectively and conveniently \cite{prieto2022wearable}. 
Wearable devices are increasingly being used to track and evaluate sleep patterns by capturing PPG signals, raw triaxis accelerometer signal (ACC), and activity-based signals like Activity (ACT). PPG measures changes in the blood volume of vascular tissues, allowing measurements of vital parameters, such as heart rate, respiration rate, arterial oxygen saturation, and blood pressure \cite{MEJIAMEJIA2022}.
On the other hand, the ACT is derived from the ACC signal processing and provides an estimation of rest and wakefulness periods to assess sleep patterns \cite{banfi2021efficient}. Besides the advantage of collecting different signals in the same device, the compact hardware of wearable devices ensures minimal discomfort and does not impact the quality of sleep. 

Sleep consists of two main stages: rapid eye movement (REM) and non-rapid eye movement (NREM). According to the guidelines provided by the American Academy of Sleep Medicine (AASM) \cite{aasm2007, aasm2012}, sleep stages can be classified into five distinct categories: (i) wakefulness (W), (ii) non-REM stage 1 (N1), (iii) non-REM stage 2 (N2), (iv) non-REM stage 3 (N3), and (v) REM stage (R). Accurately identifying and categorizing sleep stages is essential for effectively monitoring sleep patterns. However, by establishing only an accurate and reliable classification for sleep-wake stages, it is possible to obtain data to calculate important sleep quality metrics, such as: total sleep time; total wake time; sleep latency, that is the amount of time between the beginning of bedtime to actually falling asleep; sleep efficiency, calculated as total sleep time divided by total amount of time in bed; and wakefulness after sleep onset (WASO), measured as total time of wakefulness after sleep onset. \cite{shrivastava2014interpret}

While wearable devices cannot directly measure brain activity like EEG devices used in sleep stage analysis \cite{}, they can capture PPG signals, which can be utilized to estimate Heart Rate (HR) \cite{MEJIAMEJIA2022}. HR is known to decrease during the transition from wakefulness to non-REM stages \cite{silvani2008}. Additionally, there are indications that HR exhibits a slight decrease during the transition from non-REM to REM sleep stages \cite{silvani2008, habib2023}. Therefore, we hypothesize that only using the information of the HR and the Heart Rate Variability (HRV), extracted from the PPG signals, along with the ACT, derived from the ACC signals, is possible to accurately classify sleep and wake stages.


Recent works in the field of sleep-wake classification proposed to use both PPG and/or ACC signals. Most of them are based on features extracted of these signals. Table \ref{tab: comparison} displays a summary of the state-of-the-art on this topic.
\cite{fonseca2017} used a Bayesian Classifier to predict sleep-wake stages based on a set of HRV features computed from interbeat intervals obtained from PPG signals along with measures from ACC. 
Likewise, \cite{Eyal2017} used a similar approach as \cite{fonseca2017}, without the measures from ACC. 
\cite{ucar2018} proposed to use k-Nearest Neighbors (KNN) and Support Vector Machine (SVM) algorithms based on PPG and HRV features. 
\cite{palotti2019benchmark} compared a cohort of classification algorithms to perform the sleep-wake classification based on classic machine learning or deep learning techniques using only ACC signals. 
Likewise, \cite{banfi2021efficient} proposed to use only raw ACC signals using Convolutional Neural Networks (CNN) for this binary classification. 
\cite{habib2023} proposed a CNN derived from PPG raw signals of 10 subjects with sleep-disordered breathing, using a leave-one-out strategy on the sleep-wake classification with data augmentation. 
\cite{motin2023} used the same dataset as \cite{habib2023}, extracting 72 features from the PPG signals, instead of using the raw PPG, using three different classifiers, KNN, SVM and a Random Forest (RF). 

\begin{table}[!h]
\renewcommand{\arraystretch}{1.2}
\centering
\caption{\centering Summary of state-of-the-art obtained results for the sleep-wake classification.}
\label{tab: comparison}
\resizebox{\columnwidth}{!}{%
\begin{tabular}{cccccccc}
\hline
 & \textbf{Accuracy} & \textbf{Sensitivity} & \textbf{Specificity} & \textbf{F1-score} & \textbf{Kappa} & \textbf{Dataset} & \textbf{Method} \\ \hline
\cite{fonseca2017} & 91.5 $\pm$ 5.1 & 58.2 $\pm$ 17.3 & 92.9 $\pm$ 2.0 & - & 0.55 $\pm$ 0.14 & Private (s = 101) & \begin{tabular}[c]{@{}c@{}}Bayesian classifier -\\ PPG features and ACC\end{tabular} \\
\cite{Eyal2017} & 84.3 & 38.1 & 91.7 & - & 0.31 & Private (s = 88) & \begin{tabular}[c]{@{}c@{}}Bayesian classifier -\\ PPG features\end{tabular} \\
\cite{ucar2018} & 79.23 & 78.0 & 80.0 & 79.0 & 0.58 & Private (s = 10) & \begin{tabular}[c]{@{}c@{}}SVM -\\ PPG features\end{tabular} \\
\cite{ucar2018} & 79.36 & 77.0 & 81.0 & 79.0 & 0.59 & Private (s = 10) & \begin{tabular}[c]{@{}c@{}}KNN -\\ PPG features\end{tabular} \\
\cite{palotti2019benchmark} & 81.8 $\pm$ 1.0 & 90.40 $\pm$ 1.20 & 68.10 $\pm$ 1.90 & 84.30 $\pm$ 1.10 & - & MESA (s = 1,817) & \begin{tabular}[c]{@{}c@{}}Extra Trees - \\ 370 ACC features\end{tabular} \\
\cite{palotti2019benchmark} & 83.1 $\pm$ 1.0 & 91.40 $\pm$ 1.10 & 69.90 $\pm$ 2.00 & 85.50 $\pm$ 1.00 & - & MESA (s = 1,817) & \begin{tabular}[c]{@{}c@{}}LSTM 100 - \\ ACC raw signal\end{tabular} \\
\cite{motin2019sleep} & 72.36 & 70.64 & 74.22 & - & - & Private (s = 5) & \begin{tabular}[c]{@{}c@{}}Medium Gaussian SVM -\\ PPG 17 features\end{tabular} \\
\cite{motin2020} & 81.10 & 81.06 & 82.50 & 81.74 & - & Private (s = 10) & \begin{tabular}[c]{@{}c@{}}Cubic SVM -\\ PPG 22 features\end{tabular} \\
\cite{banfi2021efficient} & - & 89.20 & 92.0 & 90.9 & 0.782 & Private (s = 81) & \begin{tabular}[c]{@{}c@{}}LightCNNA -\\ ACC raw features\end{tabular} \\
\cite{habib2023} & 94.18 $\pm$ 11.95 & 94.4 & - & 93.05 $\pm$ 13.77 & 0.864 $\pm$ 0.265 & Private (s = 10) & \begin{tabular}[c]{@{}c@{}}CNN -\\ PPG raw signal\end{tabular} \\
\cite{motin2023} & 83.75 $\pm$ 0.85 & 87.79 $\pm$ 1.10 & 73.63 $\pm$ 2.45 & 80.01 $\pm$ 1.88 & - & Private (s = 10) & \begin{tabular}[c]{@{}c@{}}KNN -\\ PPG 72 features\end{tabular} \\
\cite{motin2023} & 84.66 $\pm$ 0.99 & 87.41 $\pm$ 1.24 & 77.79 $\pm$ 0.93 & 82.32 $\pm$ 0.90 & - & Private (s = 10) & \begin{tabular}[c]{@{}c@{}}SVM -\\ PPG 72 features\end{tabular} \\
\cite{motin2023} & 85.22 $\pm$ 0.62 & 87.86 $\pm$ 1.48 & 77.67 $\pm$ 3.26 & 82.45 $\pm$ 1.62 & - & Private (s = 10) & \begin{tabular}[c]{@{}c@{}}RF -\\ PPG 72 features\end{tabular} \\
 \hline
\end{tabular}%
}
\end{table}

In this work we present a method to classify sleep and wake stages, comparing the results of three commonly used machine learning techniques: Logistic Regression (LR), Random Forest (RF), and the eXtreme Gradient Boosting (XGBoost). These algorithm are based on features extracted from PPG and ACT signals with a reduced number of features for deployment feasibility on wearable devices. Additionally, a stratified analysis was conducted considering age and gender factors. 
Our method demonstrates advancement over existing approaches by reducing the feature set, thereby enabling implementation on computational-constrained wearable devices.

\section{Materials and Methods}

\begin{figure}[!h]
      \centering
      \includegraphics[width=0.95\linewidth]{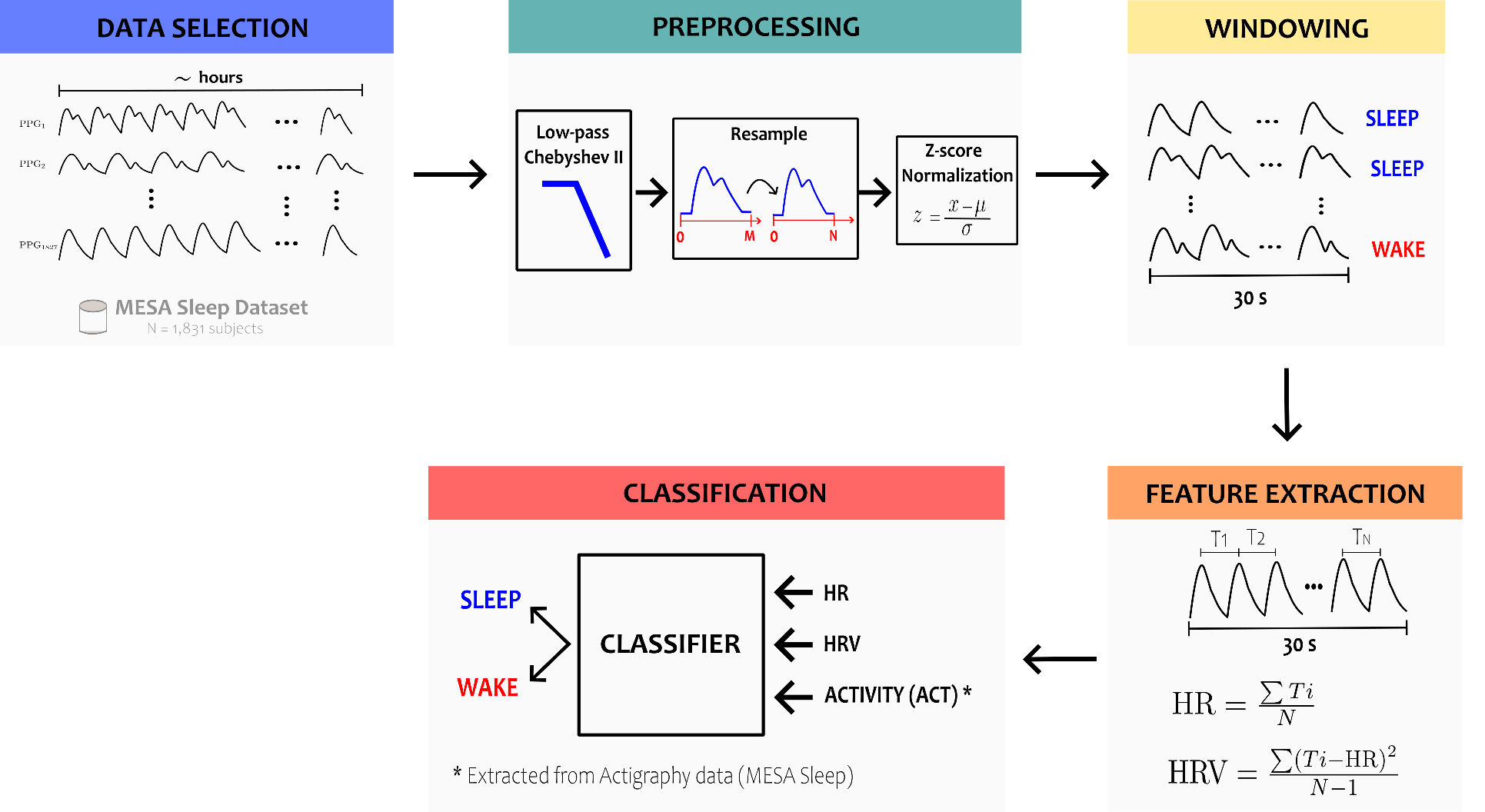}
      \caption{Overview of the proposed method for sleep-wake classification.}
      \label{fig: metodo}
   \end{figure}

The proposed approach comprises five sequential steps: Data selection, Preprocessing, Windowing, Feature Extraction, and Classification.
To begin, we describe the data utilized, specifically selecting subjects with both ACT and PPG signals. We merge the sleep stages to generate a Wake and Sleep dataset, forming the foundation for subsequent analysis.
Subsequently, we present the preprocessing and windowing procedures, detailing how the data is prepared for further processing. Additionally, we outline the feature extraction process. 
Furthermore, we describe our employed classifiers, providing information on our experimental setup.
A flow diagram summarizing these steps is shown in Fig. \ref{fig: metodo}. 

\subsection{Data Selection} 
\label{subsec:dataset}

The experiments were carried out on the MESA Sleep dataset, a subset of the Multi-Ethnic Study of Atherosclerosis (MESA) dataset \cite{zhang2018national} \cite{chen2015racial}. The MESA Sleep dataset contains data collected from 2,237 subjects, including overnight Polysomnography exams along with their corresponding PPG signals, 7-day wrist-worn Actigraphy signals, sleep stage labels for every 30 s windows, and sleep questionnaires. For this study, only patients with both ACT and PPG signals were used, resulting in 1,831 patients. The metadata information about the employed dataset is shown in Table \ref{tab:d}. 

\begin{table}[!b]
\centering
\caption{Metadata of the MESA dataset.}
\label{tab:d}
\setlength{\tabcolsep}{8pt}
\begin{tabular}{ll}
\hline
\textbf{Parameter} & \textbf{Value} \\ \hline
Subjects & $2,237$ \\
\quad \qquad with PSG & $2,056$ \\
\quad \qquad with actigraphy & $2,158$ \\
\quad \qquad with PSG and actigraphy & $1,831$ \\
 & \\
Age & $69.6 \pm 9.1$ \\
Male subjects & $1,039$ \\
\hline
\end{tabular}
\end{table}

This dataset provides sleep stage labels for every 30 s window. The labeling process followed the AASM guidelines, which suggest five sleep stage classes, as aforementioned: (i) wake (W), (ii) non-REM-1 (N1), (iii) non-REM-2 (N2), (iv) non-REM-3 (N3), and (v) REM (R) \cite{aasm2007}. Since our goal is to detect only sleep or wake stages, classes (ii), (iii), (iv), and (v) were grouped into class ``sleep'' (S). For every 30 s window, the PPG signal was sampled at 256 Hz, and the corresponding ACT registered.    

\subsection{Preprocessing}
\label{subsec:preprocessing}

We used the same preprocessing steps proposed by \cite{kotzen2022sleepppg}in this study.
The PPG data was filtered using a low-pass 8th-order Chebyshev Type II filter at 8 Hz, followed by downsampling from 256 Hz to 34 Hz using linear interpolation. Outlier values greater than or less than three standard deviations from the mean were clipped, and the data was normalized using z-score normalization. 

\subsection{Windowing}
\begin{figure}[!h]
      \centering
      \includegraphics[width=0.7\linewidth]{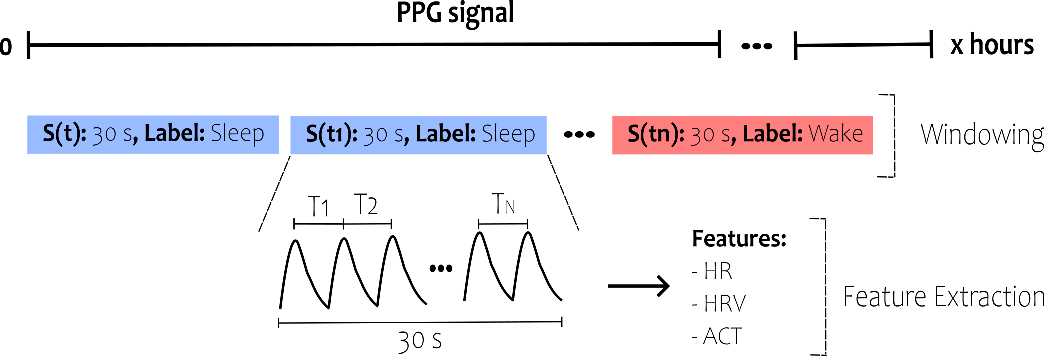}
      \caption{Windowing and Feature Extraction process example.}
      \label{fig: windowing}
\end{figure}

We partitioned the PPG signals into non-overlapping 30-second windows, as the PSG exam provides a sleep stage label for every 30 seconds of recording time.
Due to the inherent variable recording lengths for each subject in the MESA Sleep dataset, the number of windows differed among subjects. Figure \ref{fig: windowing}  illustrates the windowing extraction step applied to the PPG signal of a single subject.

\subsection{Feature Extraction}

Heart Rate (HR) and Heart Rate Variability (HRV) are widely recognized as important features associated with sleep stages and disorders, being both regulated by the sympathetic and parasympathetic nervous systems \cite{STEIN201247}. 
HR was calculated for every window by taking the mean time difference between peaks, which were detected using the method proposed in \cite{bishop2018multi}. HRV was calculated as the standard deviation of the time difference between the peaks. Windows with HR greater than 180 beats per minute or HRV outside of two standard deviations from the mean HRV of the entire dataset were discharged. As consequence, all samples from four participants were excluded in this phase, resulting in $s = 1,827$ subjects. Our model also used as input the activity value provided by the MESA-sleep acquired with the Actiware-Sleep version 5.59 analysis software (Mini-Mitter Co, Inc, Bend, OR). 

Therefore, for each $i$ subject ($i = 1, 2, \dots, s$), the features collected in each window were arranged as

\begin{eqnarray}\label{Mdata}
\mathbf{X}_i &=&
\begin{bmatrix}
x_{11} & x_{12} &  x_{13} \\
x_{21} & x_{22} &  x_{23} \\
\vdots & \vdots &  \vdots \\
x_{n1} & x_{n2} &  x_{n3}
\end{bmatrix}
,
\end{eqnarray}

where $s$ is the number of subjects, $m$ is the number of features ($m = 3$), and $n$ is the number of windows for each subject. Overall, we have a total of 2,050,280 windows: 63.93\% (1,310,690) are labeled as sleep and the remaining 36.07\% (739,590) as wake.

\subsection{Classification}
\label{subsec:classification}
Our sleep-wake classification was performed using three commonly used machine learning techniques: Logistic Regression (LR) \cite{LR}, Random Forest (RF) \cite{RF} and eXtreme Gradient Boosting (XGBoost) \cite{xgboost}.
A 10-fold cross-validation approach was employed to prevent bias in the training and testing split, with samples from the same patient grouped in the same fold to avoid intra-patient bias. The results were compared with the sleep-wake classification provided by the actigraphy used in the MESA Sleep study and with a similar study using the same dataset \cite{palotti2019benchmark}. Additionally, the results were compared with other similar studies. 


The results with the proposed approach were compared to the literature by using four different evaluation metrics commonly used in sleep-wake classification studies: Accuracy (Ac), Sensitivity (Se), Specificity (Sp), F1-score, and Cohen’s Kappa Coefficient (Kappa). These metrics were used to quantify the accuracy and agreement of the sleep-wake classification results. 

Experiments were performed using a Foxconn HPC M100-NHI with an 8-GPU cluster of NVIDIA Tesla V100 16GB cards. The model was implemented in Python (3.8.10) with the support of the libraries scikit-learn (1.1.3), XGBoost (1.6.1), and scipy (1.8.1).

\section{Results}

Table \ref{tab: results} summarizes the overall performance of the proposed methods for the sleep-wake classification task. 
Our algorithm that best performed (XGBoost) achieved Ac, Se, Spe, F1-score, and Kappa of $77.62 \pm 0.56\%$, $91.14 \pm 1.15\%$, $53.66 \pm 1.11\%$, $83.88 \pm 0.56$ and $0.480 \pm 0.008$, respectively. 
Additionally, we used the same metrics to evaluate the classification provided by the actigraphy used in the MESA Sleep dataset as baseline. It obtained Se, Spe, F1-score, and Kappa of 50.49\%, 94.75\%, 64.08\%, and 0.478, respectively.


\begin{table}[!h]
\renewcommand{\arraystretch}{1.2}
\caption{\centering Obtained results for the sleep-wake classification.}
\label{tab: results}
\centering
\begin{tabular}{lccccc}
\hline
                             & \textbf{Accuracy} & \textbf{Sensitivity} & \textbf{Specificity} & \textbf{F1-score} & \textbf{Kappa}  \\ \hline
\textit{XGBoost}             & 77.62 $\pm$ 0.56    & 91.14 $\pm$ 1.15       & 53.66 $\pm$ 1.11       & 83.88 $\pm$ 0.56    & 0.480 $\pm$ 0.008 \\
\textit{Logistic Regression} & 74.62 $\pm$ 0.66    & 96.46 $\pm$ 0.33       & 35.91 $\pm$ 0.88       & 82.92 $\pm$ 0.55    & 0.370 $\pm$ 0.009 \\
\textit{Random Forest}       & 73.80 $\pm$ 0.48    & 83.68 $\pm$ 0.93       & 56.30 $\pm$ 0.76       & 80.32 $\pm$ 0.54    & 0.413 $\pm$ 0.007 \\ \hline
\textit{MESA Actigraphy}       & -    & 50.49       & 94.75       & 64.08    & 0.478 \\ \hline
\end{tabular}
\end{table}

\begin{figure}[!h]
      \centering
      \includegraphics[width=0.6\linewidth]{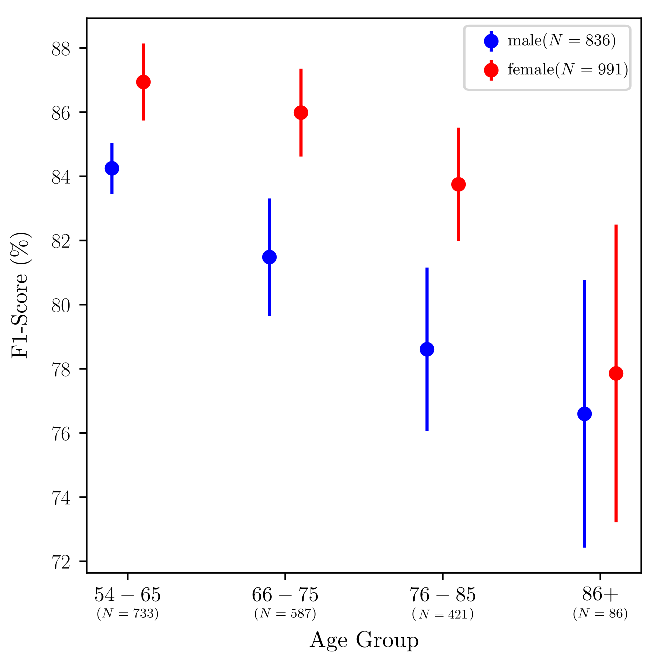}
      \caption{Stratified F1-score results for sleep-wake classification task for age and gender.}
      \label{fig:estrat}
   \end{figure}

To assess possible biases in our method, we also performed a stratified analysis of our obtained results by age and gender. In Fig. \ref{fig:estrat}, we show the obtained F1-score of our approach in four age groups (54 -- 65, 66 -- 75, 76 -- 85, and 86+) and two gender groups (male and female). 

\section{Discussion}

We developed a methodology with results comparable to the state-of-the-art while maintaining low complexity. Our approach used only three features: two obtained from the peak-to-peak interval from the photoplethysmography signal and one from the actigraph value. The peak-to-peak derived features were extracted even in adverse acquisition scenarios and the actigraph value was easily extracted from the accelerometer. This is a major advantage compared to other reports that extract a complex set of features that can be compromised in noisy acquisitions. As shown in Table \ref{tab: results}, apart from the Spe, our proposed method achieved higher metric values than the MESA Sleep actigraphy. 
This shows the gain of using PPG-derived physiological measures (HR and HRV). 

Compared to the best method from \cite{palotti2019benchmark}, which also used the MESA Sleep dataset, we obtained comparable Se ($91.14 \pm 1.15\%$ vs $91.40 \pm 1.10\%$) and F1-score ($83.88 \pm 0.56\%$ vs $85.50 \pm 1.00\%$); however, the Spe value is lower ($53.66 \pm 1.11\%$ vs $69.90 \pm 2.00\%$). 
The lower Spe might be due to the imbalance between the classes. Adopting a proper strategy for dealing with unbalanced data might improve the observed results and need further investigation. 

It is not possible to make a direct comparison with other works, since all of them are based on private datasets. 
\cite{motin2023} achieved comparable results as ours in terms of F1-score (see Table \ref{tab: comparison}), however, they didn't used ACC signals and used a much larger set of PPG features than us. Moreover, their work are based only on 10 distinct subjects. 
Even though their results are promising, this reduced number of subjects may lead to limited generalizability, along with biased learning since there is a high interdependence among intra-subject heartbeats \cite{deChazal, incor_ppg}. 
Likewise, \cite{habib2023} used a limited number of subjects, and employed a CNN on PPG raw signal, achieving the best F1-score to date, to the best of our knowledge.
On the other hand, \cite{banfi2021efficient} also achieved F1-score higher than 90\% by using only raw ACC signals, but they also have a limited number of subjects. 
However, these deep learning approaches usually requires lots of data for training, computational resources, are more complex than traditional machine learning models and often have many hyperparameters.


In the stratified analysis shown in Fig. \ref{fig:estrat}, the F1-score progressively decreases with age. Also, there was consistently better mean F1-score for females than for males, despite an increase in the standard deviation for both genders. This indicates that the reduction in the number of older patients affects both genders similarly. However, the cause of the higher mean F1-score for females cannot be determined based on the available data, as it could be due to a larger sample size or physiological differences between genders. Further research may be necessary to determine the underlying reasons for these results. 

Furthermore, it was aforementioned that deep learning models could be cumbersome for deployment in wearable devices. However, the computational resource needed depends on a set of parameters that should be investigated. Thus, future works should address this by comparing the performance and computational resources of different models. Likewise, future works should also be done on external datasets to validate the generalizability of our proposed method.

Finally, the validation of wearable sleep monitoring devices would reduce the cost of sleep disorder compared to traditional polysomnography exams. This requires  reliable and practical sleep stages prediction models  easy to implement and capable of being integrated into commercial devices in line with the reported findings. 

\section{Conclusion}
We provided evidence for a simple method to classify sleep-wake states using only three features: HR, HRV, and ACT. Our method showed comparable results with more complex methods and superior results than using only the ACT measured by an actigraph, which is currently the wearable device of choice for sleep monitoring.

\bibliographystyle{unsrtnat}
\bibliography{references}  

\begin{thebibliography}{33}
\providecommand{\natexlab}[1]{#1}
\providecommand{\url}[1]{\texttt{#1}}
\expandafter\ifx\csname urlstyle\endcsname\relax
  \providecommand{\doi}[1]{doi: #1}\else
  \providecommand{\doi}{doi: \begingroup \urlstyle{rm}\Url}\fi

\bibitem[Ramar et~al.(2021)Ramar, Malhotra, Carden, Martin, Abbasi-Feinberg,
  Aurora, Kapur, Olson, Rosen, Rowley, et~al.]{ramar2021sleep}
Kannan Ramar, Raman~K Malhotra, Kelly~A Carden, Jennifer~L Martin, Fariha
  Abbasi-Feinberg, R~Nisha Aurora, Vishesh~K Kapur, Eric~J Olson, Carol~L
  Rosen, James~A Rowley, et~al.
\newblock Sleep is essential to health: an american academy of sleep medicine
  position statement.
\newblock \emph{Journal of Clinical Sleep Medicine}, 17\penalty0 (10):\penalty0
  2115--2119, 2021.

\bibitem[Hoevenaar-Blom et~al.(2011)Hoevenaar-Blom, Spijkerman, Kromhout,
  van~den Berg, and Verschuren]{hoevenaar2011sleep}
Marieke~P Hoevenaar-Blom, Annemieke~MW Spijkerman, Daan Kromhout, Julia~F
  van~den Berg, and WM3198203 Verschuren.
\newblock Sleep duration and sleep quality in relation to 12-year
  cardiovascular disease incidence: the morgen study.
\newblock \emph{Sleep}, 34\penalty0 (11):\penalty0 1487--1492, 2011.

\bibitem[Knutson and Van~Cauter(2008)]{knutson2008associations}
Kristen~L Knutson and Eve Van~Cauter.
\newblock Associations between sleep loss and increased risk of obesity and
  diabetes.
\newblock \emph{Annals of the New York Academy of Sciences}, 1129\penalty0
  (1):\penalty0 287--304, 2008.

\bibitem[Rahe et~al.(2015)Rahe, Czira, Teismann, and
  Berger]{rahe2015associations}
Corinna Rahe, Maria~Eszter Czira, Henning Teismann, and Klaus Berger.
\newblock Associations between poor sleep quality and different measures of
  obesity.
\newblock \emph{Sleep medicine}, 16\penalty0 (10):\penalty0 1225--1228, 2015.

\bibitem[Hillman et~al.(2006)Hillman, Murphy, Antic, and
  Pezzullo]{hillman2006economic}
David~R Hillman, Anita~Scott Murphy, Ral Antic, and Lynne Pezzullo.
\newblock The economic cost of sleep disorders.
\newblock \emph{Sleep}, 29\penalty0 (3):\penalty0 299--305, 2006.

\bibitem[Streatfeild et~al.(2021)Streatfeild, Smith, Mansfield, Pezzullo, and
  Hillman]{streatfeild2021social}
Jared Streatfeild, Jackson Smith, Darren Mansfield, Lynne Pezzullo, and David
  Hillman.
\newblock The social and economic cost of sleep disorders.
\newblock \emph{Sleep}, 44\penalty0 (11):\penalty0 zsab132, 2021.

\bibitem[Krystal and Edinger(2008)]{KRYSTAL2008}
Andrew~D. Krystal and Jack~D. Edinger.
\newblock Measuring sleep quality.
\newblock \emph{Sleep Medicine}, 9:\penalty0 S10--S17, 2008.
\newblock ISSN 1389-9457.
\newblock \doi{10.1016/S1389-9457(08)70011-X}.
\newblock The Art of Good Sleep Proceedings from the 5th International Sleep
  Disorders Forum: Novel Outcome Measures of Sleep, Sleep Loss and Insomnia.

\bibitem[Rundo and Downey~III(2019)]{rundo2019polysomnography}
Jessica~Vensel Rundo and Ralph Downey~III.
\newblock Polysomnography.
\newblock \emph{Handbook of clinical neurology}, 160:\penalty0 381--392, 2019.

\bibitem[Prieto-Avalos et~al.(2022)Prieto-Avalos, Cruz-Ramos,
  Alor-Hern{\'a}ndez, S{\'a}nchez-Cervantes, Rodr{\'\i}guez-Mazahua, and
  Guarneros-Nolasco]{prieto2022wearable}
Guillermo Prieto-Avalos, Nancy~Aracely Cruz-Ramos, Giner Alor-Hern{\'a}ndez,
  Jos{\'e}~Luis S{\'a}nchez-Cervantes, Lisbeth Rodr{\'\i}guez-Mazahua, and
  Luis~Rolando Guarneros-Nolasco.
\newblock Wearable devices for physical monitoring of heart: A review.
\newblock \emph{Biosensors}, 12\penalty0 (5):\penalty0 292, 2022.

\bibitem[Mejía-Mejía et~al.(2022)Mejía-Mejía, Allen, Budidha, El-Hajj,
  Kyriacou, and Charlton]{MEJIAMEJIA2022}
Elisa Mejía-Mejía, John Allen, Karthik Budidha, Chadi El-Hajj, Panicos~A.
  Kyriacou, and Peter~H. Charlton.
\newblock 4 - photoplethysmography signal processing and synthesis.
\newblock In John Allen and Panicos Kyriacou, editors,
  \emph{Photoplethysmography}, pages 69--146. Academic Press, 2022.
\newblock ISBN 978-0-12-823374-0.
\newblock \doi{10.1016/B978-0-12-823374-0.00015-3}.

\bibitem[Banfi et~al.(2021)Banfi, Valigi, di~Galante, d’Ascanio, Ciuti, and
  Faraguna]{banfi2021efficient}
Tommaso Banfi, Nicol{\`o} Valigi, Marco di~Galante, Paola d’Ascanio, Gastone
  Ciuti, and Ugo Faraguna.
\newblock Efficient embedded sleep wake classification for open-source
  actigraphy.
\newblock \emph{Scientific reports}, 11\penalty0 (1):\penalty0 1--12, 2021.

\bibitem[Iber(2007)]{aasm2007}
Conrad Iber.
\newblock The aasm manual for the scoring of sleep and associated events:
  rules, terminology, and technical specification.
\newblock \emph{(No Title)}, 2007.

\bibitem[Berry et~al.(2012)Berry, Budhiraja, Gottlieb, Gozal, Iber, Kapur,
  Marcus, Mehra, Parthasarathy, Quan, Redline, Strohl, Davidson, and
  Tangredi]{aasm2012}
RB~Berry, R~Budhiraja, DJ~Gottlieb, D~Gozal, C~Iber, VK~Kapur, CL~Marcus,
  R~Mehra, S~Parthasarathy, SF~Quan, S~Redline, KP~Strohl, Ward~SL Davidson,
  and MM~Tangredi.
\newblock Rules for scoring respiratory events in sleep: update of the 2007
  aasm manual for the scoring of sleep and associated events. deliberations of
  the sleep apnea definitions task force of the american academy of sleep
  medicine.
\newblock \emph{J Clin Sleep Med}, 8\penalty0 (5):\penalty0 597--619, 2012.
\newblock \doi{10.5664/jcsm.2172}.

\bibitem[Shrivastava et~al.(2014)Shrivastava, Jung, Saadat, Sirohi, and
  Crewson]{shrivastava2014interpret}
Deepak Shrivastava, Syung Jung, Mohsen Saadat, Roopa Sirohi, and Keri Crewson.
\newblock How to interpret the results of a sleep study.
\newblock \emph{Journal of community hospital internal medicine perspectives},
  4\penalty0 (5):\penalty0 24983, 2014.

\bibitem[Silvani(2008)]{silvani2008}
Alessandro Silvani.
\newblock Physiological sleep-dependent changes in arterial blood pressure:
  Central autonomic commands and baroreflex control.
\newblock \emph{Clinical and Experimental Pharmacology and Physiology},
  35\penalty0 (9):\penalty0 987--994, 2008.
\newblock \doi{10.1111/j.1440-1681.2008.04985.x}.

\bibitem[Habib et~al.(2023)Habib, Motin, Penzel, Palaniswami, Yearwood, and
  Karmakar]{habib2023}
Ahsan Habib, Mohammod~Abdul Motin, Thomas Penzel, Marimuthu Palaniswami, John
  Yearwood, and Chandan Karmakar.
\newblock Performance of a convolutional neural network derived from ppg signal
  in classifying sleep stages.
\newblock \emph{IEEE Transactions on Biomedical Engineering}, 70\penalty0
  (6):\penalty0 1717--1728, 2023.
\newblock \doi{10.1109/TBME.2022.3219863}.

\bibitem[Fonseca et~al.(2017)Fonseca, Weysen, Goelema, Møst, Radha,
  Lunsingh~Scheurleer, van~den Heuvel, and Aarts]{fonseca2017}
Pedro Fonseca, Tim Weysen, Maaike~S. Goelema, Els~I.S. Møst, Mustafa Radha,
  Charlotte Lunsingh~Scheurleer, Leonie van~den Heuvel, and Ronald~M. Aarts.
\newblock {Validation of Photoplethysmography-Based Sleep Staging Compared With
  Polysomnography in Healthy Middle-Aged Adults}.
\newblock \emph{Sleep}, 40\penalty0 (7):\penalty0 zsx097, 06 2017.
\newblock \doi{10.1093/sleep/zsx097}.

\bibitem[Eyal and Baharav(2017)]{Eyal2017}
Shuli Eyal and Anda Baharav.
\newblock Sleep insights from the finger tip: How photoplethysmography can help
  quantify sleep.
\newblock In \emph{2017 Computing in Cardiology (CinC)}, pages 1--4, 2017.
\newblock \doi{10.22489/CinC.2017.274-197}.

\bibitem[Uçar et~al.(2018)Uçar, Bozkurt, Bilgin, and Polat]{ucar2018}
Muhammed~Kursad Uçar, Mehmet~Recep Bozkurt, Cahit Bilgin, and Kemal Polat.
\newblock {Automatic sleep staging in obstructive sleep apnea patients using
  photoplethysmography, heart rate variability signal and machine learning
  techniques}.
\newblock \emph{Neural Computing and Applications}, 29:\penalty0 1--16, 2018.
\newblock \doi{10.1007/s00521-016-2365-x}.

\bibitem[Palotti et~al.(2019)Palotti, Mall, Aupetit, Rueschman, Singh,
  Sathyanarayana, Taheri, and Fernandez-Luque]{palotti2019benchmark}
Joao Palotti, Raghvendra Mall, Michael Aupetit, Michael Rueschman, Meghna
  Singh, Aarti Sathyanarayana, Shahrad Taheri, and Luis Fernandez-Luque.
\newblock Benchmark on a large cohort for sleep-wake classification with
  machine learning techniques.
\newblock \emph{NPJ digital medicine}, 2\penalty0 (1):\penalty0 1--9, 2019.

\bibitem[Motin et~al.(2023)Motin, Karmakar, Palaniswami, Penzel, and
  Kumar]{motin2023}
Mohammod~Abdul Motin, Chandan Karmakar, Marimuthu Palaniswami, Thomas Penzel,
  and Dinesh Kumar.
\newblock Multi-stage sleep classification using photoplethysmographic sensor.
\newblock \emph{Royal Society Open Science}, 10\penalty0 (4):\penalty0 221517,
  2023.
\newblock \doi{10.1098/rsos.221517}.

\bibitem[Motin et~al.(2019)Motin, Karmakar, Penzel, and
  Palaniswami]{motin2019sleep}
Mohammod~Abdul Motin, Chandan~Kumar Karmakar, Thomas Penzel, and Marimuthu
  Palaniswami.
\newblock Sleep-wake classification using statistical features extracted from
  photoplethysmographic signals.
\newblock In \emph{2019 41st Annual International Conference of the IEEE
  Engineering in Medicine and Biology Society (EMBC)}, pages 5564--5567. IEEE,
  2019.

\bibitem[Motin et~al.(2020)Motin, Karmakar, Palaniswami, and Penzel]{motin2020}
Mohammod~Abdul Motin, Chandan Karmakar, Marimuthu Palaniswami, and Thomas
  Penzel.
\newblock {Photoplethysmographic-based automated sleep–wake classification
  using a support vector machine}.
\newblock \emph{Physiol. Meas.}, 41:\penalty0 075013, 2020.
\newblock \doi{10.1088/1361-6579/ab9482}.

\bibitem[Zhang et~al.(2018)Zhang, Cui, Mueller, Tao, Kim, Rueschman, Mariani,
  Mobley, and Redline]{zhang2018national}
Guo-Qiang Zhang, Licong Cui, Remo Mueller, Shiqiang Tao, Matthew Kim, Michael
  Rueschman, Sara Mariani, Daniel Mobley, and Susan Redline.
\newblock The national sleep research resource: towards a sleep data commons.
\newblock \emph{Journal of the American Medical Informatics Association},
  25\penalty0 (10):\penalty0 1351--1358, 2018.

\bibitem[Chen et~al.(2015)Chen, Wang, Zee, Lutsey, Javaheri, Alc{\'a}ntara,
  Jackson, Williams, and Redline]{chen2015racial}
Xiaoli Chen, Rui Wang, Phyllis Zee, Pamela~L Lutsey, Sogol Javaheri, Carmela
  Alc{\'a}ntara, Chandra~L Jackson, Michelle~A Williams, and Susan Redline.
\newblock Racial/ethnic differences in sleep disturbances: the multi-ethnic
  study of atherosclerosis (mesa).
\newblock \emph{Sleep}, 38\penalty0 (6):\penalty0 877--888, 2015.

\bibitem[Kotzen et~al.(2022)Kotzen, Charlton, Salabi, Amar, Landesberg, and
  Behar]{kotzen2022sleepppg}
Kevin Kotzen, Peter~H Charlton, Sharon Salabi, Lea Amar, Amir Landesberg, and
  Joachim~A Behar.
\newblock Sleepppg-net: a deep learning algorithm for robust sleep staging from
  continuous photoplethysmography.
\newblock \emph{IEEE Journal of Biomedical and Health Informatics}, 2022.

\bibitem[Stein and Pu(2012)]{STEIN201247}
Phyllis~K. Stein and Yachuan Pu.
\newblock Heart rate variability, sleep and sleep disorders.
\newblock \emph{Sleep Medicine Reviews}, 16\penalty0 (1):\penalty0 47--66,
  2012.
\newblock ISSN 1087-0792.
\newblock \doi{10.1016/j.smrv.2011.02.005}.

\bibitem[Bishop and Ercole(2018)]{bishop2018multi}
Steven~M Bishop and Ari Ercole.
\newblock Multi-scale peak and trough detection optimised for periodic and
  quasi-periodic neuroscience data.
\newblock In \emph{Intracranial Pressure \& Neuromonitoring XVI}, pages
  189--195. Springer, 2018.

\bibitem[Stoltzfus(2011)]{LR}
Jill~C. Stoltzfus.
\newblock Logistic regression: A brief primer.
\newblock \emph{Academic Emergency Medicine}, 18\penalty0 (10):\penalty0
  1099--1104, 2011.
\newblock \doi{10.1111/j.1553-2712.2011.01185.x}.

\bibitem[Breiman(2001)]{RF}
Leo Breiman.
\newblock Random forests.
\newblock \emph{Machine Learning}, 45:\penalty0 5--32, 2001.
\newblock \doi{10.1023/A:1010933404324}.

\bibitem[Chen and Guestrin(2016)]{xgboost}
Tianqi Chen and Carlos Guestrin.
\newblock {XGBoost}.
\newblock In \emph{Proceedings of the 22nd {ACM} {SIGKDD} International
  Conference on Knowledge Discovery and Data Mining}. {ACM}, aug 2016.
\newblock \doi{10.1145/2939672.2939785}.

\bibitem[de~Chazal et~al.(2004)de~Chazal, O'Dwyer, and Reilly]{deChazal}
Philip de~Chazal, M.~O'Dwyer, and R.B. Reilly.
\newblock Automatic classification of heartbeats using ecg morphology and
  heartbeat interval features.
\newblock \emph{IEEE Transactions on Biomedical Engineering}, 51\penalty0
  (7):\penalty0 1196--1206, 2004.
\newblock \doi{10.1109/TBME.2004.827359}.

\bibitem[Costa et~al.(2023)Costa, Dias, Cardenas, Toledo, Lima, Krieger, and
  Gutierrez]{incor_ppg}
Thiago Bulhões Da~Silva Costa, Felipe~Meneguitti Dias, Diego Armando~Cardona
  Cardenas, Marcelo Arruda Fiuza~De Toledo, Daniel Mário~De Lima, Jose~Eduardo
  Krieger, and Marco~Antonio Gutierrez.
\newblock Blood pressure estimation from photoplethysmography by considering
  intra- and inter-subject variabilities: Guidelines for a fair assessment.
\newblock \emph{IEEE Access}, 11:\penalty0 57934--57950, 2023.
\newblock \doi{10.1109/ACCESS.2023.3284458}.

\end{thebibliography}






\end{document}